# Title: Potentials and Limitations of Large-scale, Individual-level Mobile Location Data for Food Acquisition Analysis

Duanya Lyu [a], Luyu Liu [b,]*, Catherine Campbell [c], Yuxuan Zhang [d], Xiang Yan [a]

[a] *Department of Civil & Coastal Engineering, University of Florida, Gainesville FL, United States*
[b] *Department of Geosciences, Auburn University, Auburn AL, United States*
[c] *Department of Family, Youth and Community Sciences, University of Florida, Gainesville FL, United States*
[d] *Department of Computer & Information Sciences & Engineering, University of Florida, Gainesville FL, United States*

* Corresponding author: Luyu Liu, luyuliu@auburn.edu
2046J Haley Center, Auburn University, Auburn, AL 36849, United States

**Abstract**

Understanding food acquisition is crucial for developing strategies to combat food insecurity, a major public health concern. The emergence of large-scale mobile location data (typically exemplified by GPS data), which captures people's movement over time at high spatiotemporal resolutions, offer a new approach to study this topic. This paper evaluates the potential and limitations of large-scale GPS data for food acquisition analysis through a case study. Using a high-resolution dataset of 286 million GPS records from individuals in Jacksonville, Florida, we conduct a case study to assess the strengths of GPS data in capturing spatiotemporal patterns of food outlet visits while also discussing key limitations, such as potential data biases and algorithmic uncertainties. Our findings confirm that GPS data can generate valuable insights about food acquisition behavior but may significantly underestimate visitation frequency to food outlets. Robustness checks highlight how algorithmic choices-especially regarding food outlet classification and visit identification-can influence research results. Our research underscores the value of GPS data in place-based health studies while emphasizing the need for careful consideration of data coverage, representativeness, algorithmic choices, and the broader implications of study findings.

# 1. Introduction

*Food insecurity,* which refers to the lack of stable access to sufficient, safe, and nutritious food for a healthy, active life (Simelane & Worth, 2020), is a major public health concern (Caspi et al., 2012). In 2023, 13.5% of U.S. households experienced food insecurity (Matthew P. Rabbitt et al., 2025). A critical aspect of addressing this issue is understanding food acquisition, i.e., the processes and actions through which people obtain food. Examining how households acquire food from food-selling places to their home provides essential insights for developing effective interventions and policies to combat food insecurity (Rabbitt et al., 2023).

Prior studies have commonly used surveys to gather detailed data on household food acquisition behavior, including shopping habits, spending, and food consumption (Coleman-Jensen et al., 2019). Well-structured surveys enable comparisons of food acquisition behavior across groups and longitudinal analyses (Anekwe & Zeballos, 2019). Also, interviews and focus groups provide qualitative insights such as how cultural and socio-economic factors influence food acquisition (Shier et al., 2022). However, these methods usually rely on personal memories to report the spatiotemporal information associated with food store visits, which may result in bias, inaccuracies, and limited spatial and temporal insights (Hillier et al., 2017). In recent years, researchers have used Global Positioning System (GPS) devices to augment traditional approaches, such as conducting geo-tagged surveys (Elliston et al., 2020) as well as distributing tracking devices to record geo-fenced visits (Wray et al., 2023) or mobility trajectories (Zenk et al., 2011). By tracking people's movement with high spatial and temporal resolution (Chen et al., 2016), GPS data allows one to reconstruct their activity-travel pattern and support more detailed spatiotemporal analyses. For instance, researchers can examine food exposure based on activity spaces, beyond one's home or workplace (Elliston et al., 2020). However, collecting GPS-augmented survey data is resource-intensive and so only small sample sizes are typically achieved, making it challenging to engage representative participant groups. Cetateanu and Jones (2016) and Siddiqui *et al.* (2024) reviewed the papers on GPS and food environment exposure and reported sample sizes ranging from 12 to 654 individuals. Moreover, the awareness of carrying monitoring devices can also lead to increased consciousness of actions and potential behavioral changes over the study period (Zhang et al., 2021).

With the widespread use of location-enabled mobile devices such as smartphones and smart watches, large-scale GPS data collected by location-tracking apps are increasingly common (Smith et al., 2023; Xie et al., 2023; Zhou et al., 2022). These datasets, made available to the research community by data vendors such as SafeGraph, Cuebiq, and Gravy Analytics, capture the movements of millions of individuals over extended periods (e.g., weeks, months, and years), with spatial and temporal resolutions in meters and seconds. Such data offer a new approach to study human mobility (Kwan, 2016; K. Zhao et al., 2016). For example, they have been used to analyze neighborhood-, region- or even national-level associations between food outlet visits and related health and social outcomes (Chang et al., 2022; Hu et al., 2021; Xie et al., 2023; R. Xu et al., 2023). So far, most published works analyze these data at the aggregate levels such as census tracts or points-of-interest (POIs), partly because SafeGraph made its aggregate-level mobile location data freely available to researchers around the world from 2020 to 2023 (Xie et al., 2023; K. Zhao et al., 2016). While such data provide valuable insights on human mobility at the aggregate level, they do not allow one to examine behavioral differences at the individual level. Also, since these data are pre-processed by vendors using proprietary algorithms, the research community has limited insight into potential biases in both the derived mobility metrics (e.g., food outlet visits) and the underlying algorithms. Despite their large sample sizes and high resolution, GPS datasets often have uneven spatial and temporal coverage and can exhibit sample bias, such as the underrepresentation of disadvantaged groups (Li et al., 2023). Moreover, other researchers have noted that analytical choices in processing GPS data—such as algorithm design and parameter setting—can significantly impact the reliability and accuracy of the derived mobility measures (Kwan, 2016).

In light of the above discussions, this paper analyzes a unique dataset—GPS mobility traces from millions of individuals in Florida—to explore the potentials and limitations of using large-scale, longitudinal, individual-level GPS data for food acquisition analysis. We first present the analytical steps involved to process raw GPS data collected from smart mobile devices into food-acquisition-related metrics for further analysis. By breaking down the algorithmic considerations and parameter choices made in each step, we reveal how algorithmic uncertainties in GPS data processing can influence the results. Subsequently, using a case study of Jacksonville, Florida, we discuss the advantages of leveraging large-scale GPS data for food acquisition analysis compared to traditional methods, such as surveys, as well as the limitations posed by potential data biases and algorithmic uncertainties. Specifically, we identify the novel insights GPS data can provide by examining spatiotemporal patterns of food acquisition behavior at the individual level. We also explore the potential limitations by examining data representativeness and evaluating the robustness of study findings to several key parameter choices.

## 2. Processing GPS Data for Food Acquisition Analysis

To obtain food acquisition metrics from raw mobile location data (i.e., GPS data) for further analysis requires extensive data processing. Figure 1 shows the key analytical steps we have employed here, where many algorithmic considerations are informed by the existing literature. Specifically, the process begins with processing the mobile phone GPS data to infer each device user's home location and activities (or stays) so as to obtain their activity-travel patterns. Next, we integrate GPS data and food outlet location data to infer individual food acquisition behavior. Two key considerations here are regarding 1) which food outlets to be included for food acquisition analysis and 2) the proximity criteria used to assign an observed activity (or stay) close to a food outlet as a "food acquisition visit." Based on the inferred food acquisition visits, we can then calculate food acquisition metrics and analyze their spatiotemporal patterns. We now discuss each key step in detail.

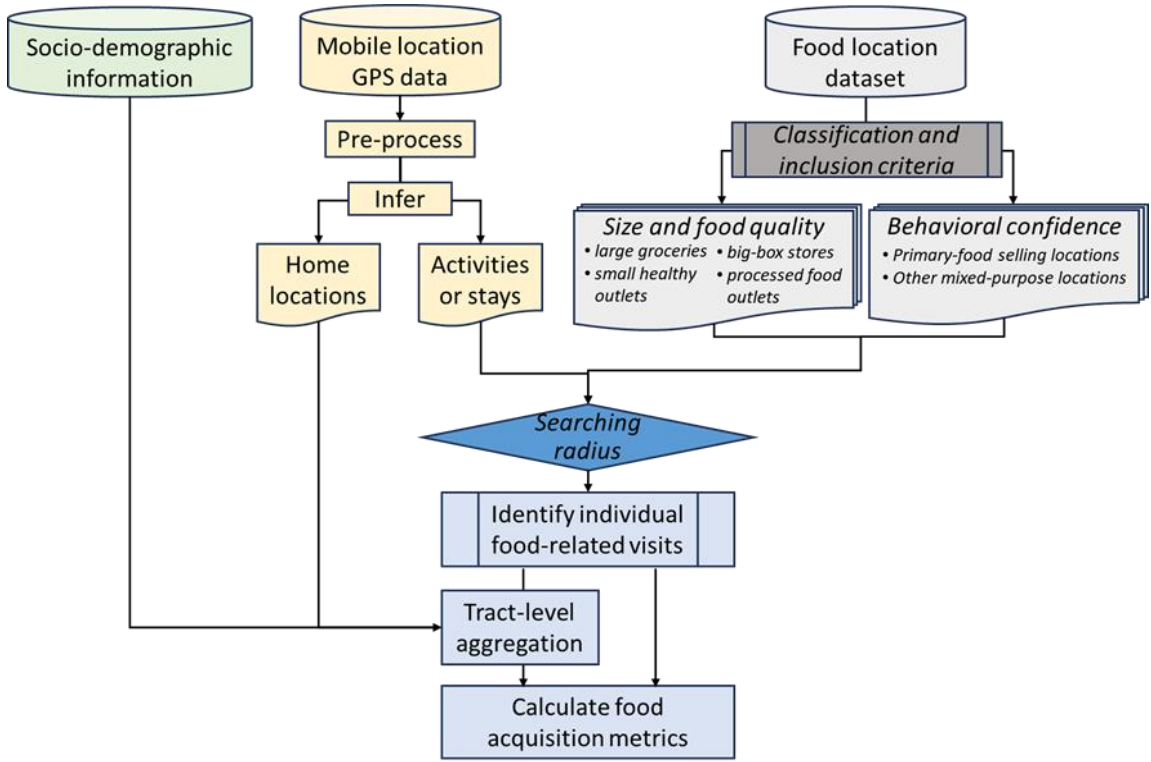

Figure 1. Analytical steps to obtain food acquisition metrics from GPS data

*2.1. Home Location and Activity Stay Inference*

Considering that home location is crucial for analyzing food acquisition behavior (Coleman-Jensen et al., 2019; Zenk et al., 2018), we inferred home location for device users. In this study, we use the proxy-home-location algorithm developed in (X. Zhao et al., 2022). This approach segments the study area into 20-meter grids, counts GPS points recorded between 10:00 PM and 6:00 AM within each grid, and designates the grid with the highest GPS point density as the home location. For users whose home locations could not be inferred due to a lack of nighttime GPS points (approximately 16.2% of device users), we leveraged weekend data. Assuming users typically spend most of their daytime at home on weekends, we used GPS points recorded between 6:00 AM and 10:00 PM to estimate their home locations.

Moreover, we extracted activity stays from the GPS data using the *Trackintel* package, which employs a time-space heuristic method. This method identifies activity stays (referred to as *stops* in the package) as periods of minimal movement and detects them using a sliding window algorithm that clusters points (Martin et al., 2023). In this study, we set thresholds for stop detection to a 100-meter radius and a duration between 5 and 720 minutes. To reduce outliers, we excluded activity stays exceeding two hours, aligning with the American Time Use Survey (ATUS), which reports a median grocery shopping duration of 30 minutes with a standard deviation of 30.6 minutes (Brown & Borisova, 2007). Additionally, we extracted the origins of those activity stays using the package's backward searching method, which clusters spatiotemporally linked GPS points.

*2.2. Identifying Food Outlet Visits*

The activity stays derived from GPS data do not contain information about activity purposes, which is typically inferred based on the proximity of an activity location to adjacent places of interest (SafeGraph, n.d.). When identifying food outlet visits from GPS data, a key consideration is determining which food-selling stores to include. One should also classify food outlets, which can vary significantly in the type, price, quantity, and variety of food they offer. Prior research on food acquisition has shown that regional food environments consist of a wide variety of food outlets, each influencing community health in different ways and exhibiting distinct visitation patterns (Balagtas et al., 2023; Shier et al., 2022; Todd & Scharadin, 2016). Another complicating factor involves individuals also visiting different food outlets for multiple purposes; for example, some visit big box stores for non-food items and gas stations for food. USDA's FoodAPS has indicated that Supplemental Nutrition Assistance Program (SNAP) households allocate 13% of their food spending to convenience stores, dollar stores, and pharmacies (Todd & Scharadin, 2016).

Here we developed a two-dimensional classification approach to categorize food outlets (Todd & Scharadin, 2016; R. Xu et al., 2023). First, food locations were grouped into four groups based on by size and the quality: *large groceries/supermarkets* that primarily sell food (e.g., Trader Joe's and Aldi); *big-box stores* that offer a full range of food products along with other goods (e.g., Walmart Supercenter); *small healthy food outlets* that carry healthy items include supplements and medicines (i.e. drug stores) and grocery items like fresh produce, diary and eggs (e.g., food marts, dollar stores)[1]; and *processed food outlets* that sell only processed and low-nutrition food (e.g., gas station stores and food store with processed food only). The second dimension classified locations by the confidence of food acquisition when visiting them: outlets primarily selling food where individuals predominantly visit for food acquisition (e.g., groceries, food marts) versus those visited for various other purposes (e.g., big-box stores, gas stations).

*2.3. Food Acquisition Metrics*

We then integrated the food outlet dataset and activity patterns derived from GPS data to infer food outlet visits. Specifically, we identified food outlet visits from activity stays using a buffer-based approach, assigning predefined radii based on outlet types: 50 meters for Small Healthy Outlets and Processed Food Outlets, 200 meters for Big-box Stores, and 150 meters for Large Groceries.[2] Based on the inferred food outlet visits, we further calculated four metrics widely discussed in the food access literature to assess food

---

[1] In this study, we classify dollar stores as small food outlets that carry healthy items, despite their typical association with unhealthy food options and their potential contribution to food deserts by limiting the entry of supermarkets (Chenarides et al., 2021a), for the following reasons: First, increasing research highlights the role of dollar stores in enhancing food accessibility and security by providing affordable, widely available produce and other healthy options (John et al., 2023). These stores may help fill gaps in food access, particularly in areas where other retailers are unwilling to operate (Chenarides et al., 2021a). Furthermore, an increasing number of dollar stores now carry fresh produce and dairy products. Ongoing initiatives and implementations by dollar store companies have highlighted their commitment to expanding fresh produce offerings. For example, as of Q1 2023, Dollar General has introduced fruits and *(continue)* vegetables in nearly 3,900 of their locations (Dollar General Corporation, n.d.). Notably, we differentiate gas station stores from dollar stores, as the former primarily offer processed or low-quality food.

[2] Section 3.4 provides additional details to explain these radius choices.

acquisition patterns (Leroy et al., 2015; Todd & Scharadin, 2016).

First, we calculated the number of food outlet visits made by each individual during the study period, mirroring survey-based approaches that measure food acquisition frequency. To capture diversity in food acquisition, we counted the number of unique stores visited by each user. This metric resembles traditional surveys that assess the diversity of people's food sources. Regarding spatial metrics used in traditional methods to assess food accessibility, we computed the home-to-store distance, defined as the network distance (based on Open Street Map data) between the user's inferred home location and visited food outlets. We calculated the average distance from a user's home to all food outlets they visited, as well as the distance to the nearest food store from their home. Lastly, we calculate the proportion of home-based food outlet visits[3], a metric widely used in prior work to examine food purchases in relation to home location so as to understand the connection between residential food environments and food acquisition behaviors (Thornton et al., 2017; Ver Ploeg et al., 2015).

These metrics would allow us to assess how well GPS data can replicate and enhance traditional survey approaches for food acquisition analysis. Specifically, we will analyze visitation patterns of various food outlets using GPS data to assess their alignment with traditional survey findings and reliability. Further exploring the temporal trends (e.g., time-of-day and day-of-week patterns) and spatial aspects (e.g., home-to-store distance and its distribution) allows us to explore novel insights generated by GPS data not easily captured in traditional survey studies.

*2.4. Algorithmic Uncertainties*

The above discussion suggests that, when using GPS data for food acquisition analysis, one must make many algorithmic choices in data processing. These choices—often made based on the literature findings or the analyst's subjective judgment—may introduce significant uncertainties to study results. Notably, prior work on leveraging large-scale GPS data for human mobility analysis highlights two key challenges affecting activity detection and classification (Kwan, 2016): (1) determining *the location relevant to* the activity and (2) inferring *the activities performed* at those locations. Here we discuss the algorithmic uncertainties associated with these two challenges for food acquisition analysis.

*(1) Uncertainty in food outlet visit identification*. Radius is a crucial parameter in visit attribution, as it determines whether an activity inferred from the GPS data is considered a visit *to* a food outlet. A smaller radius risks missing visits (false negatives), while a larger radius may capture unrelated visits (false positives). For example, supermarkets may have visits missed with a smaller radius (Figure 2, left), while small grocery stores may capture unrelated visits with a larger one (Figure 2, right). In this study, we test radii of 50m, 100m, 150m, and 200m to evaluate how this parameter affects the robustness of our GPS-based food acquisition analysis.

---

[3] Here we defined home-based food outlet visits as those originating within 200 meters of the inferred home location. The 200-meter radius—as opposed to the residential parcel—is used to mitigate the potential impacts of GPS location errors (up to 100 meters).

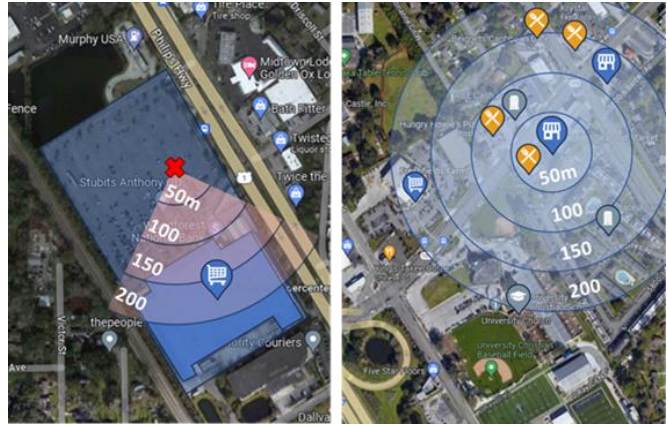

Figure 2. Identification radii in food-related trip extraction

*(2) Uncertainty in Activity Purpose Determination.* Visits to the same location may serve different purposes. In the context of food acquisition, visits to food outlets may not always be for food acquisition. To explore this issue, we considered two food outlet inclusion criteria to calculate food acquisition metrics: (a) including all locations and (b) limiting analysis to primary food-selling locations. For instance, for visits to small healthy food outlets, the former include dollar stores and drug stores but latter does not. Excluding non-primary-food-selling locations offers a more conservative estimate, potentially omitting some food acquisition visits but increasing the likelihood that the identified visits accurately represent actual food acquisition.

## 3. Case Study: GPS-data-based Food Acquisition Analysis in Jacksonville, Florida

We further conduct a case study in Jacksonville, Florida to shed light on the potentials and limitations of using large-scale, individual-level GPS data for food acquisition analysis. After describing the case study area and data sources, we investigate the sampling rate and inferred food acquisition visit to assess data coverage and representativeness. Next, we examine key food acquisition metrics derived from GPS data and the associated spatiotemporal patterns to evaluate to what extent GPS data can replicate traditional survey approaches for food acquisition analysis and generate novel insights. Finally, we conduct robustness checks to assess how algorithmic uncertainties in food outlet visit identification and activity purpose determination shape the results of GPS-data-based food acquisition analysis.

### 3.1. Case Study Area and Data Sources

Our case study area is the City of Jacksonville, Florida, the largest municipality in the state. The city's demographics reveal a complex socioeconomic landscape. According to the American Community Survey (ACS) 2018-2022 five-year estimates (U.S. Census Bureau, 2022), the population of over 950,000 includes 53.1% White, 30.4% Black, and 11.3% Hispanic or Latino, and with a median age of 36.3 years, a median household income of $64,138, and 14.8% living in poverty. Figure 3 illustrates the spatial distribution of sociodemographic characteristics by census tracts. As shown in the figure, urban tracts generally exhibit higher population density, greater percentage of individuals aged 18-39, lower household vehicle ownership, and higher poverty rates. The northwest part of the study area demonstrates a lower percentage of the White population.

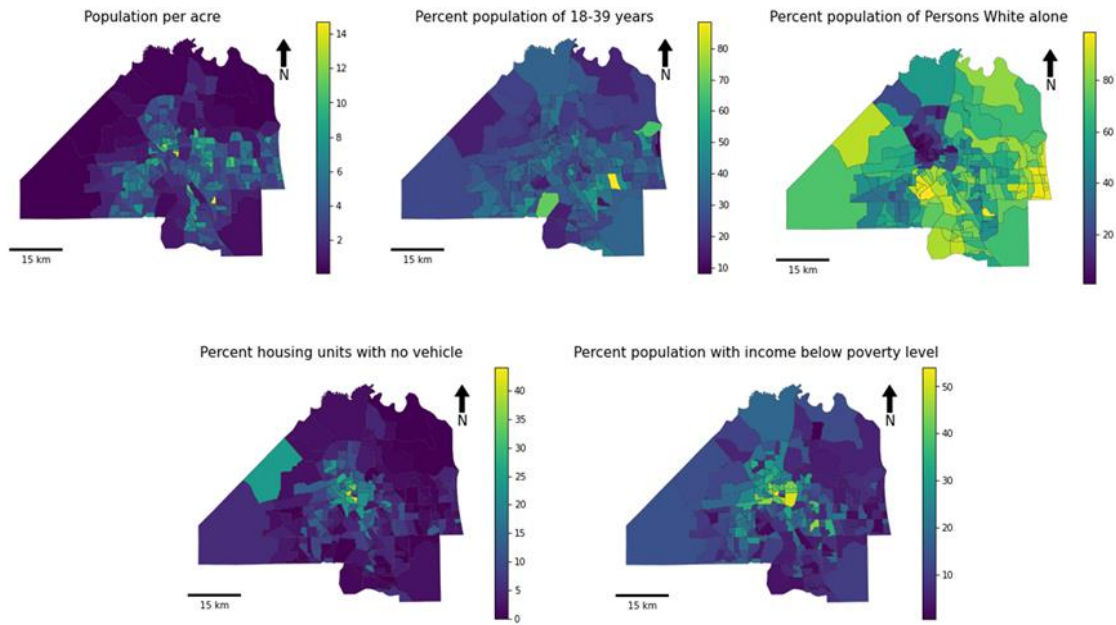

Figure 3. Tract-level socio-demographic characteristics in Jacksonville, Florida

This study uses a large-scale mobile device location dataset from Gravy Analytics, which aggregates data from over 150 million U.S. mobile devices through various apps (Gravy Analytics, 2023b). The data vendor suggests that it complies with privacy laws, sourcing data only from users who opt in, with a 48- to 72-hour processing delay (Gravy Analytics, 2023a). The dataset is also pre-processed for positioning errors, with accuracy indicated by a forensic identifier field (Gravy Analytics, 2023c; Y. Xu et al., 2022). For this study, we included only records classified as "high accuracy," where GPS positioning errors do not exceed 35 meters. This pre-analysis filtering minimizes positioning errors, ensuring more reliable results. After pre-processing, we retained 286.4 million disaggregated records from September 1 to October 15, 2022 (45 days). The data fields included device identifiers, latitude, longitude, geohash, and timestamp.

Another key data source is a comprehensive food outlet database for North Florida developed by the University of Florida GeoPlan Center, covering various components of the local food system, including food production, retail, and distribution sites (Alachua County, 2022). Our study focuses on food retail outlets that provide food-at-home (FAH) access, including grocery stores, supermarkets, drug stores, corner stores, gas station convenience stores. Figure 4 shows the distribution of the four food outlet types in Jacksonville, based on the classification in Section 2.2.

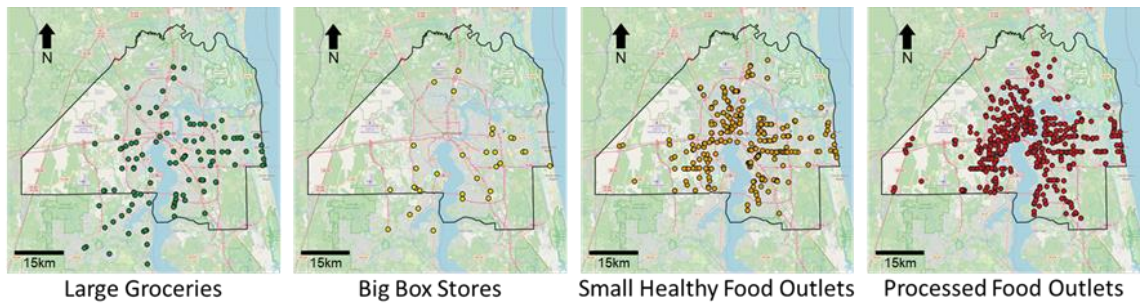

Figure 4. Distribution of various types of food outlets in the study area

## 3.2. Data Coverage and Representativeness

### 3.2.1. Sampling Rate

Figure 5 shows the spatial distribution of the 93,854 individual device home locations and the histogram of the sampling rate across census tracts. On average, 10.4% of each tract's population, based on census estimates, is represented in our sample, highlighting that GPS data achieves a better coverage than survey. Regarding the spatial distribution of sampling rates across census tracts, the map suggests a reasonable spatial coverage and the histogram shows a normal-like distribution centered around 8%. However, the two figures also show that sampling rate varies significantly across census tracts, which indicate potential spatial bias.

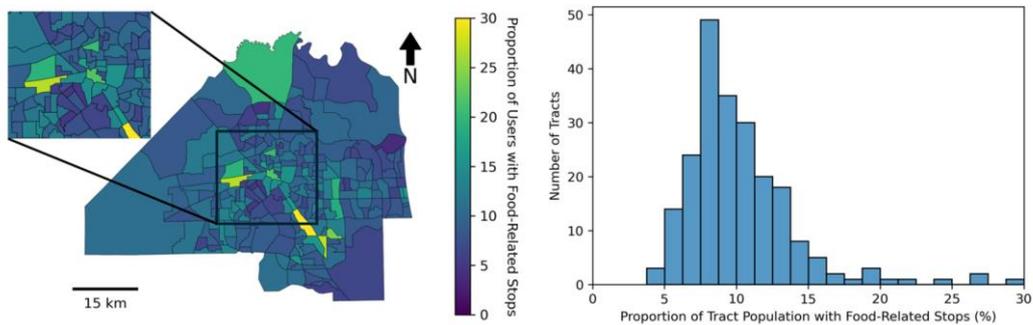

Figure 5. Distribution of sampling rate across census tracts

### 3.2.2. Inferred Food Outlet Visits

We extracted a total of 852,224 food outlet visits, with the following distribution: 250,916 for Large Groceries, 76,979 for Big Box Stores, 191,796 for Small Healthy Food Outlets, and 332,533 for Processed Food Outlets. The number of visits captured is significantly larger compared to traditional surveys [4].

## 3.3. Understanding Food Acquisition Patterns with GPS Data

### 3.3.1. Food Acquisition Metrics

In this subsection, we analyze various food acquisition metrics to examine whether the results derived from GPS data align with or differ from those obtained from surveys. Based on the inferred food outlet visits, we analyzed visitation frequency, unique stores visited, home-to-store distances, and percentage of home-based visits. Table 1 presents the population-averaged metrics derived from our GPS data.

Table 1. Food acquisition metrics for each type of store

| Metrics | Large Groceries | Big Box Stores | Small Healthy Food Outlets | Processed Food Outlets | All Food Outlets |
|---|---|---|---|---|---|

---

[4] For the "proportion of home-based visits" metric, we considered only visits with inferred origins, resulting in 1,336 visits for Large Groceries, 646 for Big Box Stores, 801 for Small Healthy Food Outlets, and 1,808 for Processed Food Outlets. The significant reduction in number of visits is because inferring the origin of these visits requires continuous GPS tracking of the trip trajectory, which is not available for most visits.

| | | | | | | |
|---|---|---|---|---|---|---|
| Number of visits per individual in 1.5 months (visits) | | 4.74 | 3.12 | 4.1 | 5.02 | 9.08 |
| Number of unique stores visited per individual in 1.5 months (stores) | | 1.86 | 1.31 | 1.93 | 2.38 | 3.85 |
| Distance of visited store to home (km) | Euclidean | 5.29 | 6.54 | 5.63 | 5.8 | 5.62 |
| | Network | 7.43 | 8.69 | 7.21 | 7.47 | 7.61 |
| Proportion of home-based visits (%) | | 18.65 | 14.44 | 18.33 | 16.02 | 17.84 |

The average number of food outlet visits per individual over 1.5 months is 9, equating to 1.4 visits per week. In contrast, USDA FoodAPS reports higher rates for the South US: 2.86, 1.33, and 2.37 *food acquisition events* per week at *Large grocery stores*, *Small and specialty stores* and *All other food-at-home stores,* respectively—equating to 18.4, 8.5, and 15.2 over 1.5 months (Todd & Scharadin, 2016). A Florida survey with 1,594 respondents reports more comparable values, finding that the most common *grocery store shopping* frequency was weekly (Hodges & Stevens, 2013). We shall should note that the referenced studies are household-based, whereas our analysis relies on individual devices, which may contribute to underestimation (Ver Ploeg et al., 2015). Moreover, we observed a significant underestimation in the proportion of home-based visits, as ATUS reports 64% of grocery shopping trips begin and end at home (Ver Ploeg et al., 2009). A main source for these underestimations is the gaps in GPS location tracking, which leaves out some activities and trips taken by the device owner. On the other hand, the number of unique stores visited is closer to the 3.5 retail banners visited per month reported in a national sample of 2,091 grocery shoppers (FMI and the Hartman Group, 2022). Moreover, the GPS data's home-to-store distance aligns closely with the literature, showing an average of 3-4 miles (4.8-6.4 km) (Euclidean distance, to primary store) (Todd & Scharadin, 2016).

The findings echo previous literature, emphasizing that reliable and valid measurement is essential when using mobility data to study food environments (Zenk et al., 2018). Inconsistent tracking may lead to missing records, resulting in the underestimation of food acquisition activity and a low proportion of home-based visits, which are familiar environments to individuals. Despite these limitations, GPS data show potential into capturing actual locations visited that were tracked and distances travelled, which are not always well-captured in self-reported surveys. While not fully replicating survey methods, GPS data complements food access research by offering more accurate insights into individuals' actual movements.

### 3.3.2. Spatial and Temporal Patterns

Mobile GPS location data continuously tracks millions of people's movements with high spatial resolution, enabling analysts to derive novel spatial and temporal insights that enhance the understanding of food acquisition behavior. This is a key advantage that GPS data has compared to traditional surveys, which usually lack the sample size, spatial and temporal coverage, or granularity for detailed spatiotemporal analyses. To empirically test and validate this point, we analyze the spatial patterns of home-to-store distances and the temporal variation in food outlet visitation. These aspects of food acquisition are important yet remain underexplored in survey-based studies due to smaller sample sizes and lower spatial or temporal resolution (Todd & Scharadin, 2016; Ver Ploeg et al., 2015).

*(1) Spatial Patterns*

Figure 6 shows the distribution curves of the distance from individual home to store. The left-side figure, depicting the home-to-nearest-store distance, reveals distinct distributions by food outlet type. These distance distributions reflect the densities of various types of food outlets, where a longer distance indicates a smaller density. Interestingly, the right-side figure, showing home-to-visited-store distances, reveals similar distance distributions across outlet types, suggesting that individuals often bypass the nearest store to shop at stores further from home (J. L. Liu et al., 2015).

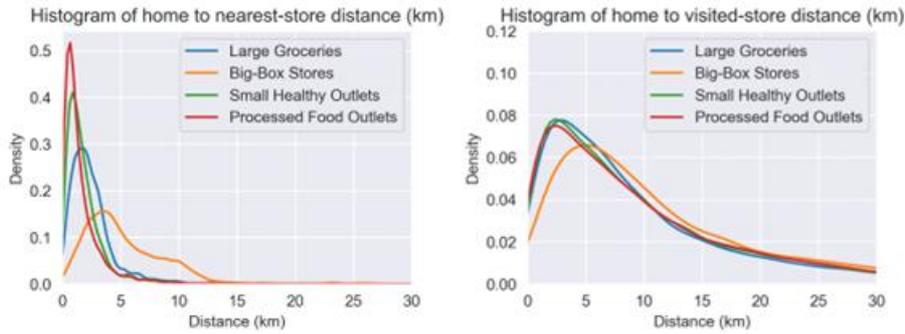

Figure 6. Distribution of home-to-store distances within the population

Figure 7 presents scatter density plots, with nearest- and visited-store distances on the x and y axes, respectively. Darker colors indicate higher density. The slopes in the figure indicate the degree to which people prefer closer stores: a slope of 1 means exclusive visits to the nearest store, while a slope of 0 means distance is not a factor. Big-box Stores show the steepest contours, indicating a strong preference for nearby locations, followed by Large Groceries. In contrast, Small Healthy Outlets and Processed Food Outlets have flatter contours, suggesting distance is less influential in choosing these stores. This pattern likely results from business strategies that standardize Big-box Stores and Large Groceries, making location differences less significant, while the diversity in quality and price at Small Healthy Outlets and Processed Food Outlets drives people to visit specific locations.

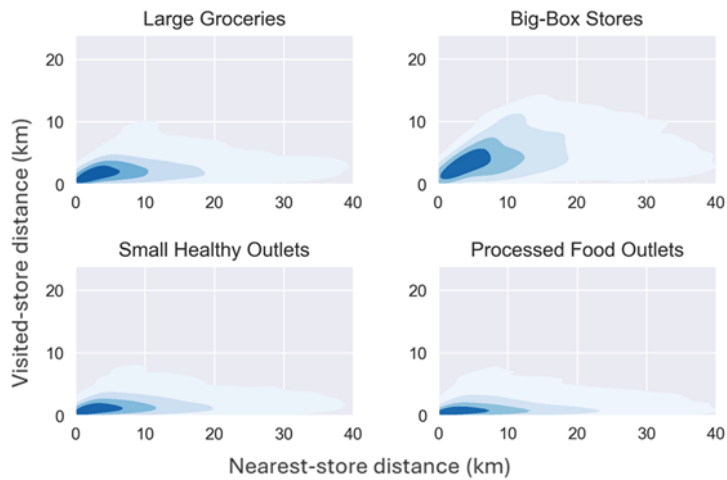

Figure 7. Density contours of home-to-store distances

We aggregated individual-level measurements to the tract level in Figure 8, mapping home-to-nearest store distance, home-to-visited store distance, and their difference. Darker colors indicate larger distances. The first row shows greater nearest-store distances in rural areas, reflecting limited food access. However, the second row reveals large visited-store distances in some urban areas, aligning with lower-income, high-density, and predominantly non-white populations, consistent with Jacksonville's urban food deserts phenomenon (Lewis et al., 2018). The last row highlights where individuals bypass closer stores. Big-box stores and large groceries generally show smaller differences, consistent with Figure 7. However, in the central-west area, Big-box Store deviations are greater, corresponding to higher-income, predominantly white populations, possibly reflecting limited nearby options or store preferences. In contrast, larger deviations for Large Groceries visits appear in the northwest, where more underprivileged populations reside, likely due to affordability or cultural preferences. Small Healthy Food and Processed Food outlets exhibit larger differences but with less clear spatial patterns.

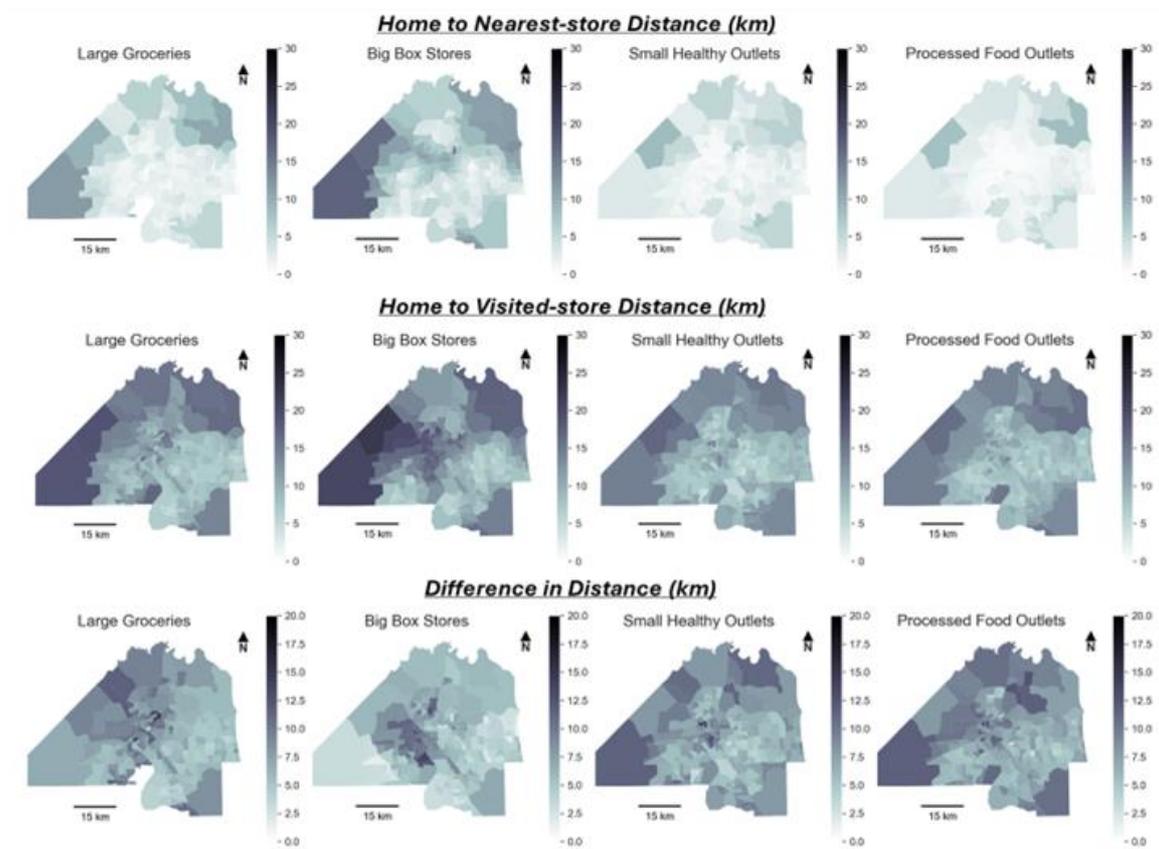

Figure 8. Spatial distribution of home-to-store distances

*(2) Temporal Pattern*

Prior research on food acquisition has shown distinct temporal visitation patterns across food outlets. For example, fast food visits peak on weekdays, while supermarkets see more traffic on weekends (East et al., 1994; García Bulle Bueno et al., 2024). Building on these findings, we leveraged GPS data to analyze temporal patterns in food outlet visitation across different temporal dimensions, including day of the week, time of day, and daily variations.

Figure 9 presents the patterns. The time-of-day curves (first row) reveal generally similar weekday and weekend patterns. Though, weekday visits vary more by outlet type with the two large outlet types showing clearer daytime peaks; weekend patterns are more uniform (except for Big-box Stores), with a weaker evening peak but a stronger midday peak, consistent with literature on weekend activity shifts (East et al., 1994). The day-of-week trends (second row) highlight a Friday peak in general and a higher weekend share for Big-box Stores, trends also noted in prior studies (Cai, 2006). The last row presents the longitudinal trends in number of visits extracted each day. It shows overall stability with a notable drop around Labor Day (September 5, 2022), suggesting a temporary shift in food shopping behavior due to the holiday.

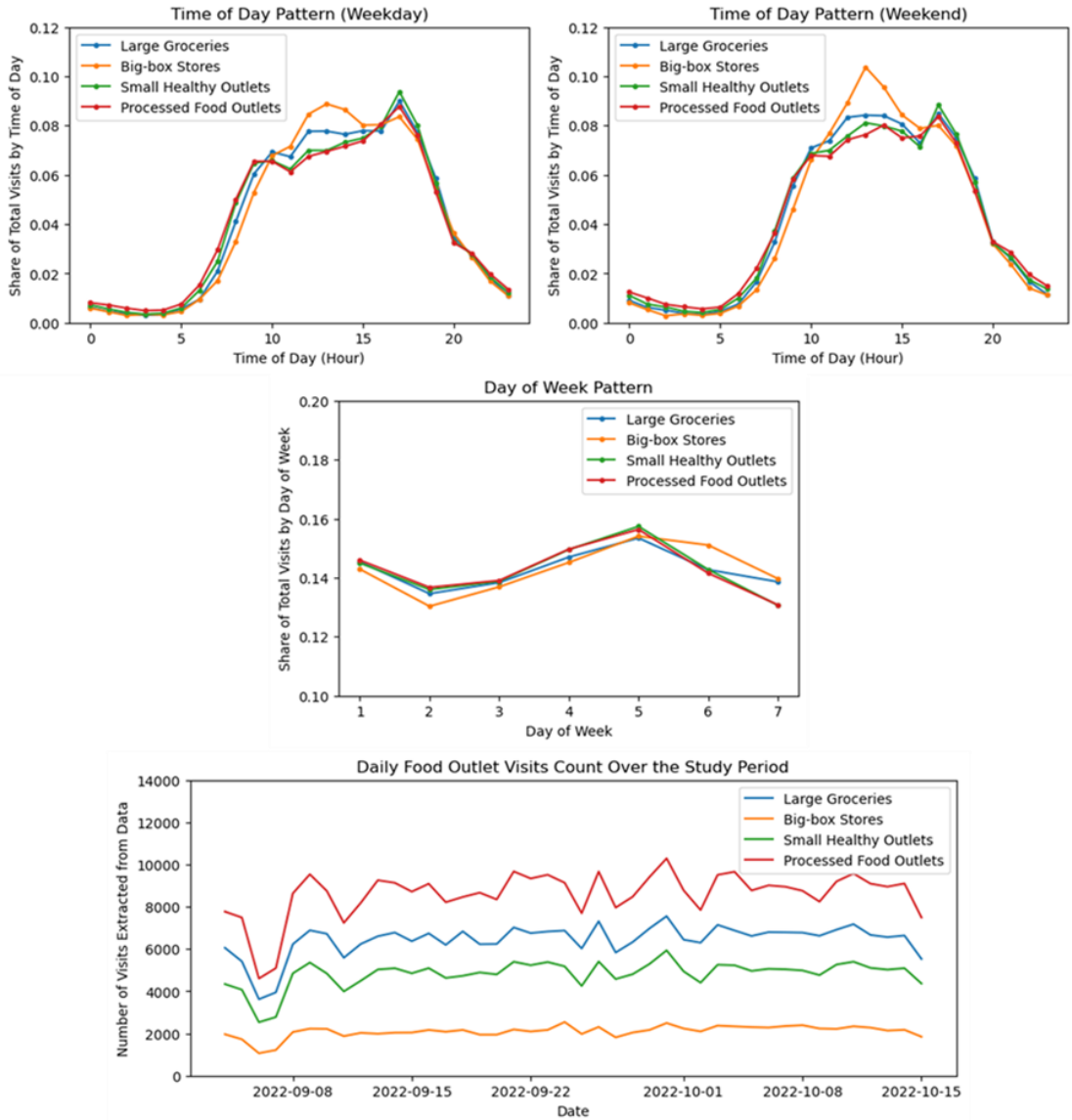

Figure 9. Temporal patterns of food acquisition activities

### 3.4. Robustness Checks

In Section 2.4, we have discussed that algorithmic uncertainties can pose a major threat to the validity and reliability of study findings on food acquisition behavior derived from GPS data. This section presents results from the robustness checks we have performed to address the two key challenges associated with leveraging GPS data for food acquisition analysis: uncertainty in food outlet visit identification and uncertainty in activity purpose determination.

*3.4.1. Uncertainty in Food Outlet Visit Identification*

Figure 10 shows the metrics calculated with food outlet visits identified under different radii. The general trend remains consistent across radius variations, with no drastic changes in patterns. However, the number of visits shows a notable increase for Small Healthy Outlets and Processed Food Outlets as the radius expands from 50m to 100m, while the increase for Big-box Stores is more gradual. This supports the expectation that food outlet visit identification is sensitive to radius selection, especially for smaller outlets, which have a limited venue area and are often clustered with other points of interest (Figure 4). These findings informed our parameter choices for the case study, that is, we selected a 50m for the two smaller outlet types, 200m for Big-box Stores, and 150m for Large Groceries.

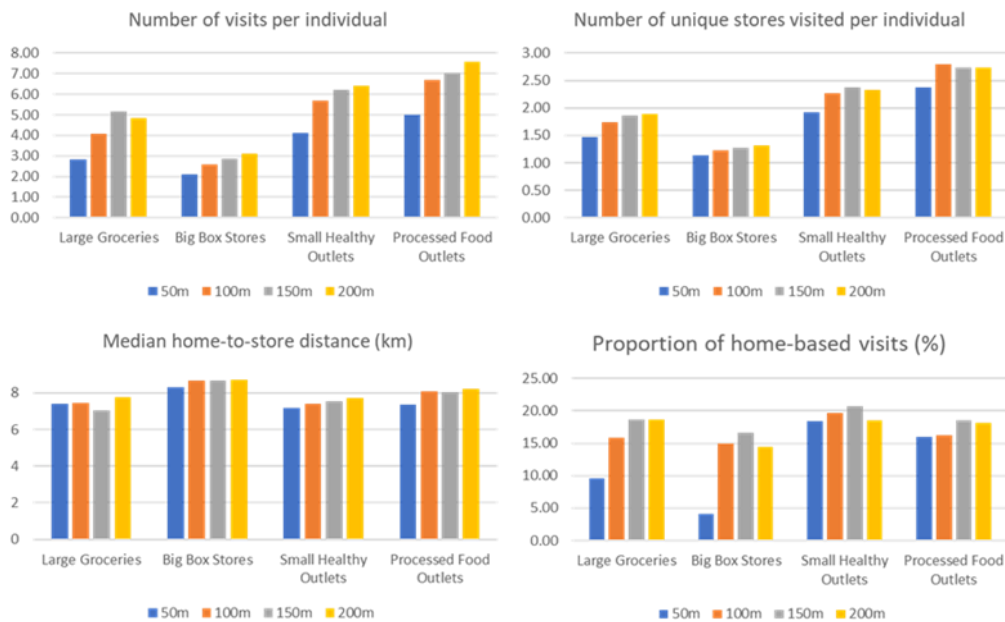

Figure 10. Food access metrics calculated under different radii

*3.4.2. Uncertainty in Activity Purpose Determination.*

Table 2 and Figure 11 present the food acquisition metrics derived from GPS data when we focus solely on primary food-selling locations. The results differ from those show in Table 2 and Figure 6 in that visits to *Big-box Stores* are now excluded.

As expected, we observed a decrease in total visits and the number of unique stores visited compared to the results shown in Table 2. Focusing on primary food-selling locations also led to smaller home-to-store distances and more home-based visits, aligning with findings from activity-space studies which shown maintenance activities to have smaller radii (Gong et al., 2020). Differentiating store types revealed distinct patterns. Visits to Processed Food Outlets showed similar trends, while visits to Large Groceries saw increased home-to-store distances, reduced home-based visits, and more frequent trips to more unique stores. These differences likely arise from the behaviors of individuals who visit exclusively food-selling stores versus those who shop at outlets with broader inventories. Shoppers at specialized grocery stores visit more often and travel farther, while those at gas stations or dollar stores tend to visit more frequently than fast-food or specialized processed outlets, consistent with findings in the literature (Todd & Scharadin, 2016; Ver Ploeg et al., 2015).

Table 2 Food acquisition metrics by food outlet type (primary food-selling locations only)

| Metrics | Large Groceries | Small Healthy Food Outlets | Processed Food Outlets | All Food Outlets |
|---|---|---|---|---|
| Number of visits per individual in 1.5 months (visits) | 5.13 | 3.47 | 6.04 | 5.13 |
| Number of unique stores visited per individual in 1.5 months (stores) | 1.87 | 1.46 | 2.41 | 1.87 |
| Distance of visited store to home (km) | 8.15 | 7.82 | 7.03 | 7.52 |
| Proportion of home-based visits (%) | 17.95 | 21.62 | 18.9 | 17.95 |

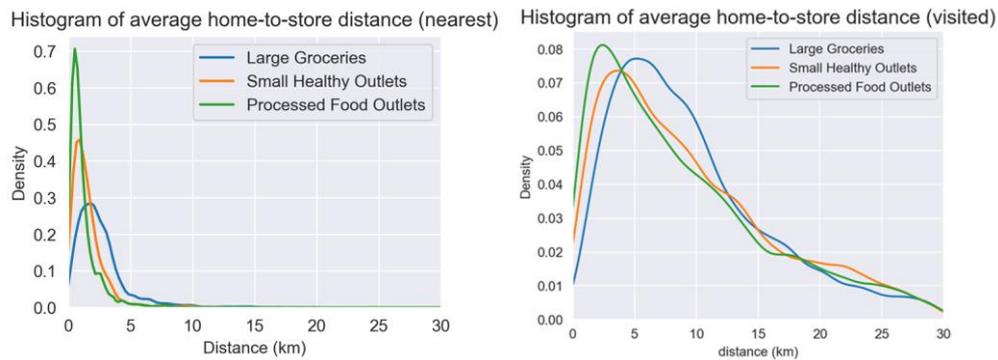

Figure 11. Distribution of home-to-store distances (primary food-selling locations only)

Overall, the robustness checks suggest that the food outlet visit identification and activity purpose determination methods are sensitive to algorithmic uncertainties. Radius selection and outlet classification can influence the number of visits, store type differentiation, and distances traveled. Despite these uncertainties, the methods still provide valuable insights into food access patterns. Adjustments to algorithmic parameters can allow for more accurate representation of food acquisition behavior, enhancing the reliability of study findings.

## 4. Discussion

### 4.1. The Potential of Using GPS Data for Food Acquisition Analysis

Our case study highlights the potential of using GPS data to significantly enhance food acquisition analysis. GPS data can be an effective tool for studying food acquisition, offering advantages over traditional methods.

First, our study demonstrates the capability of GPS data in capturing food acquisition patterns at the individual level with a large sample size, which is often challenging with traditional methods. GPS enables the calculation of individual food acquisition metrics as well as their spatial and temporal distribution across large areas and extended periods, offering a more comprehensive understanding than conventional surveys. This broad yet high-resolution coverage is particularly valuable for informing policy design and evaluation. For example, policymakers can monitor food security programs' effectiveness spatially across areas and assess their impact temporally by tracking changes in visitation patterns to food outlets before and after program implementation. This allows for more efficient adjustments to policies based on long-term trends.

Additionally, GPS data offer flexibility in defining and refining analytical approaches to suit different research questions with the same dataset. For instance, by adjusting classification criteria, we examined food acquisition patterns both with and without non-traditional food outlets. The results showed that including gas stations and dollar stores increased visits per individual from 3.47 to 5.02 and unique stores visited from 1.46 to 2.41. This flexibility can provide new angles for assessing interventions, such as the impact of introducing healthy produce into dollar stores on food accessibility and security—an important topic in food security research (Chenarides et al., 2021b; John et al., 2023; Lucan et al., 2018).

*4.2. GPS Data's Limitations for Studying Food Acquisition*

Our analysis also reveals some key limitations that influence the validity and reliability of study findings derived from the use of GPS data in food acquisition research.

First, despite large sample sizes, our GPS data exhibit spatial sampling biases, with varying rates across areas. This imbalance can contribute to ecological fallacy, where aggregate trends misrepresent individual behaviors (Chen et al., 2016). In food acquisition analysis, this is particularly problematic, as it affects the equity of food access interventions. Marginalized communities face disproportionate barriers to healthy food (Jin et al., 2023), prior studies show GPS data tend to underrepresent them (Coston et al., 2021; Li et al., 2023; Squire, 2019). Future research should address these biases and exploring the role of sociodemographic factors to improve understanding (Singleton et al., 2023).

Another issue is inconsistent tracking. From the case study, we found that food outlet visitation frequencies derived from GPS data are lower than those reported in surveys. These discrepancies could be affected by whether and when a user activates location-tracking for their mobile devices and other factors such as signal coverage and strength across space. Regarding food acquisition, many individuals may not use navigation apps for routine grocery trips, especially those originated from their home, which may result in underestimations of food outlet visits. This introduces another source of bias in GPS-based food acquisition analysis, potentially compromising both the internal and external validity of the study findings.

Additionally, the robustness checks we have performed in this study affirm that assumptions, algorithmic choices and parameter settings used in the process of inferring food acquisition metrics from GPS data can significantly influence study results (Kwan, 2016). Although we found that the general patterns and trends were consistent across the decision space, algorithmic uncertainties clearly had a major influence on both the content and reliability of food acquisition result results. As big data such as mobile location data are increasingly used for human mobility analysis, our work stresses the importance of paying attention to the algorithms used throughout the analytical process.

### *4.3. Study Limitations and Future Research*

A key limitation of this study is that we have attempted to shed light on the potentials and limitations of using GPS data for food acquisition analysis based on a single case study. The specific GPS dataset used here may not be representative of other GPS datasets used in the research community. Also, as briefly mentioned in the results section, GPS data tracks individuals rather than households, whereas food shopping is often a household-level activity, with multiple individuals from the same household visiting food outlets together (Todd & Scharadin, 2016). Our work can thus be enhanced by performing a household-level analysis. Moreover, since the case study is based on Jacksonville, FL, its findings may be influenced by the unique characteristics of the study area, potentially limiting the external validity of the results. Due to resource constraints, we could not access multiple popular GPS datasets. Future research can build on our work by analyzing multiple GPS datasets across diverse study areas to enhance validity and generalizability.

In addition, there can be concerns about the temporal and spatial generalizability of the empirical findings. Temporally, research on food sales has shown seasonality in food demand (Balagtas et al., 2023; Hu et al., 2021), which may influence mobility patterns. Therefore, generalizing our findings from the 45-day study period could introduce bias and limit the representativeness of the results. Spatially, study findings from Jacksonville may not be transferable to other contexts. A 2012 study noted disparities in food acquisition among Health Zones within the city, with Urban Core residents facing a greater health burden (Healthy Jacksonville Children Obesity Prevention Coalition, 2012). Additionally, Jacksonville's poverty rate (14.8%) exceeds both the national (12.5%) and state averages (12.9%) (U.S. Census Bureau, 2022). These socio-economic factors should be considered when generalizing the results to other contexts. Future research can apply the analytical steps outlined in Section 2 to explore food acquisition patterns across different locations and time periods, which can validate our empirical findings.

### 5. Conclusion

Mobile location data provides a novel approach to studying human mobility. This study presents a systematic analysis of the potentials and limitations of using large-scale, individual level GPS data for food acquisition analysis. Using a large-scale mobile location dataset with 286 million GPS records, we conducted a case study in Jacksonville, Florida. We inferred several food acquisition metrics commonly examined in the literature and explored their spatial and temporal patterns. The results demonstrate the capability of GPS data to extract key insights regarding food acquisition patterns that confirm findings from prior survey-based studies. On the other hand, our analysis suggests that relying on GPS data would significantly underestimate food outlet visitation frequency. The robustness checks, focusing on examining how algorithmic uncertainties in the classifications of food-selling stores and identification radii of food outlet visits shape research results, affirm general patterns and trends across the decision space but also reveal some inherent challenges of extracting food acquisition visits from GPS data.

Overall, this study confirms the potential of GPS data for analyzing food acquisition while also underscoring the need for careful interpretation and application. Our research highlights the need for critical reflexivity with respect to data coverage and representativeness, algorithmic choices, and the findings generated from them. We suggest future research to apply a mixed-methods research design and integrate diverse data sources to gain a more holistic understanding of people's food acquisition patterns. Studies integrating small-scale GPS data with surveys have uncovered novel insights that challenged previously held behavioral assumptions (B. Liu et al., 2020; Sadler et al., 2016). Similarly conducting surveys on the same population from which large-scale GPS data are collected can allow for the triangulation of results and enhance the generalizability of study findings, ultimately contributing to more effective interventions and policies for combating food insecurity.

**Author contributions**

Conceptualization: XY, CC, LY
Data Curation: YZ, DL
Formal Analysis: DL
Methodology: XY, CC, LY
Supervision: XY, CC, LY
Visualization: DL
Writing – Original Draft: DL, LL
Writing – Review and Editing: XY, CC, LY, DL

**Declarations of Interest statement**

The authors report there are no competing interests to declare.

**Funding**

This work was supported the US Department of Transportation Center for Transit-Oriented Communities Tier-1 University Transportation Center (Grant No. 69A3552348337) and the National Science Foundation (Award # 2416202).

**Acknowledgements**

We thank Erik Finlay at the University of Florida GeoPlan Center for sharing the North Florida food retailer location database with us.

**Data statement**

The data that support the findings of this study are available from Gravy Analytics. Restrictions apply to the availability of these data, which were used under license for this study. Aggregate-level data are available from the authors with the permission of Gravy Analytics. All the codes are available at https://anonymous.4open.science/r/GPS-Food-Accessibility-9B32.

Purchase Survey (FoodAPS). *International Journal of Environmental Research and Public Health*, *14*(10), 1133. https://doi.org/DOI: 10.3390/ijerph14101133

Hodges, A. W., & Stevens, T. J. (2013). Local food systems in Florida: Consumer characteristics and economic impacts. *Proceedings of the Florida State Horticultural Society*, *126*, 338–345.

Horn, A. L., Bell, B. M., Bulle Bueno, B. G., Bahrami, M., Bozkaya, B., Cui, Y., Wilson, J. P., Pentland, A., Moro, E., & De La Haye, K. (2023). Population mobility data provides meaningful indicators of fast food intake and diet-related diseases in diverse populations. *Npj Digital Medicine*, *6*(1), 208. https://doi.org/10.1038/s41746-023-00949-x

Hu, Y., Quigley, B. M., & Taylor, D. (2021). Human mobility data and machine learning reveal geographic differences in alcohol sales and alcohol outlet visits across U.S. states during COVID-19. *PLOS ONE*, *16*(12), e0255757. https://doi.org/10.1371/journal.pone.0255757

Jin, A., Chen, X., Huang, X., Li, Z., Caspi, C. E., & Xu, R. (2023). Selective Daily Mobility Bias in the Community Food Environment: Case Study of Greater Hartford, Connecticut. *Nutrients*, *15*(2). https://doi.org/10.3390/nu15020404

John, S., Sundermeir, S. M., & Gardner, K. (2023, October). *Stretching the dollar: Community-informed opportunities to improve healthy food access through dollar stores*. https://www.cspinet.org/sites/default/files/2023-10/CSPI_dollarStoreReport_2023_4.pdf

Kwan, M.-P. (2016). Algorithmic Geographies: Big Data, Algorithmic Uncertainty, and the Production of Geographic Knowledge. *Annals of the American Association of Geographers*, *106*(2), 274–282. https://doi.org/10.1080/00045608.2015.1117937
22